\documentclass[conference]{IEEEtran}
\IEEEoverridecommandlockouts
\usepackage{cite}
\usepackage{amsmath,amssymb,amsfonts}
\usepackage{algorithmic}
\usepackage{graphicx}
\usepackage{textcomp}
\usepackage{xcolor}
\usepackage{stfloats}
\usepackage{color}
\usepackage{multirow}
\usepackage{bm}
\def\BibTeX{{\rm B\kern-.05em{\sc i\kern-.025em b}\kern-.08em
    T\kern-.1667em\lower.7ex\hbox{E}\kern-.125emX}}
\begin{document}

\title{Enhanced Eye Diagram Estimation Method for Nonlinear Systems With Input Jitter}

\author{

\IEEEauthorblockN{1\textsuperscript{st} Hanqing Zhang}
\IEEEauthorblockA{\textit{College of Information Science and Electronic Engineering}\\
\textit{Zhejiang University}\\Hangzhou, Zhejiang, China\\
hanqing.zhang@zju.edu.cn}

\and

\IEEEauthorblockN{2\textsuperscript{nd} Feijun Zheng}
\IEEEauthorblockA{\textit{School of Micro-Nano Electronics} \\
\textit{Zhejiang University}\\Hangzhou, Zhejiang, China\\
zhengfj@zju.edu.cn}

}

\maketitle

\begin{abstract}
An enhanced multiple-edge response (MER) based eye diagram estimation method is proposed to evaluate the performance of nonlinear systems with input jitter. Compared with \textcolor{black}{existing MER-based methods} \textcolor{black}{which only took into account the bit effect}, the proposed method first determines both \textcolor{black}{orders of bit effect and jitter effect}.
\color{black}
These decided orders can affirm the necessary MERs.
\color{black}
\textcolor{black}{Subsequently}, the proposed method figures out \textcolor{black}{the minimal} number of sampling points so that the \textcolor{black}{necessary} MERs can be \textcolor{black}{recovered} quickly based on the Nyquist theory and can be used to \textcolor{black}{create} eye diagrams.
\color{black}
Lastly, the eye diagrams and their parameters are compared with those generated by traditional transient simulation and an existing MER-based method which introduces input jitter through a convolution process.
\color{black}
The result \textcolor{black}{indicates} that this enhanced method is more accurate than the \textcolor{black}{existing} MER-based method.
\end{abstract}


\section{Introduction}\label{I}

In the design of high-speed transmission systems, the challenge of signal integrity (SI) needs to be considered seriously. To evaluate the quality of signals, eye diagrams are used frequently. Nowadays, there are three major methods to form eye diagrams: transient simulation, worst-case eye diagram estimation, and statistical eye diagram estimation. As \textcolor{black}{the} bit error rate (BER) \textcolor{black}{becomes} lower and lower, transient simulation is unsuitable, which wastes a lot of time. But the other two methods are more effective. To implement these two methods, some techniques have been invented: single-bit response (SBR) \cite{b1}, double-edge response (DER) \cite{b2}, and multiple-edge response (MER) \cite{b3,b4}. In linear time-invariant (LTI) transmission systems, SBR and DER-based methods are effective. However, if systems are nonlinear, these two methods are not accurate enough but MER-based \textcolor{black}{methods}. In MER-based \textcolor{black}{methods}, system responses affected by previous bits are generated by \textcolor{black}{short-term} transient \textcolor{black}{simulations lasting only a few signal periods and these responses can form eye diagrams.}

\begin{figure}[t]
\centerline{\includegraphics[width=230pt]{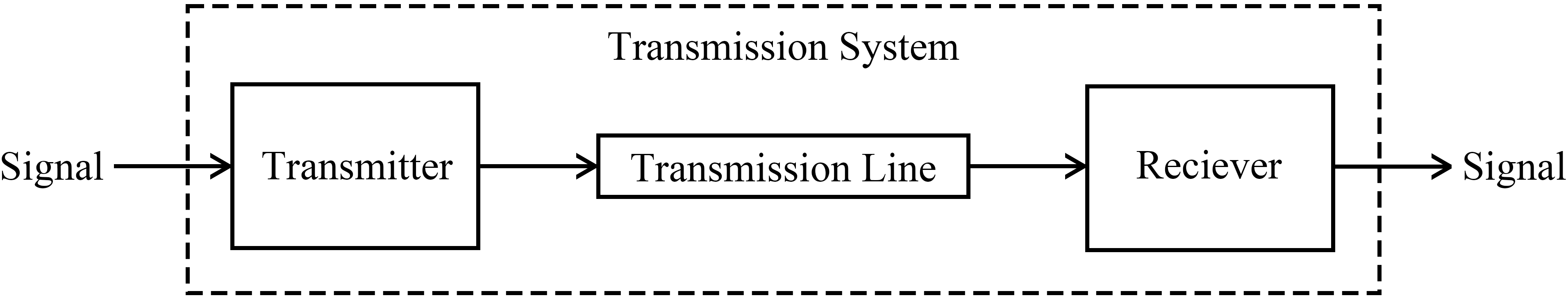}}
\caption{\textcolor{black}{Structure of transmission systems.}}
\label{Fig1}
\end{figure}

A transmission system can be represented in Fig.~\ref{Fig1}. Thus, output signals are affected by two factors which are the transmission system and input signals. Up to now, some MER-based methods have been invented to meet \textcolor{black}{the} requirements of \textcolor{black}{introducing} jitter \cite{b5,b6,b7}. However, only a few articles have claimed how to consider the jitter of input signals in nonlinear \textcolor{black}{systems}.
\color{black}
In \cite{b4} and \cite{b8}, the authors use the convolution process to introduce the jitter of input signals while estimating eye diagrams. These methods are fast and accurate in many cases. However, each estimation method has cases where it is not applicable. Especially, if there's a large degree of nonlinearity, the convolution process will give an imprecise result.
\color{black}

\begin{figure}[t]
\centerline{\includegraphics[width=210pt]{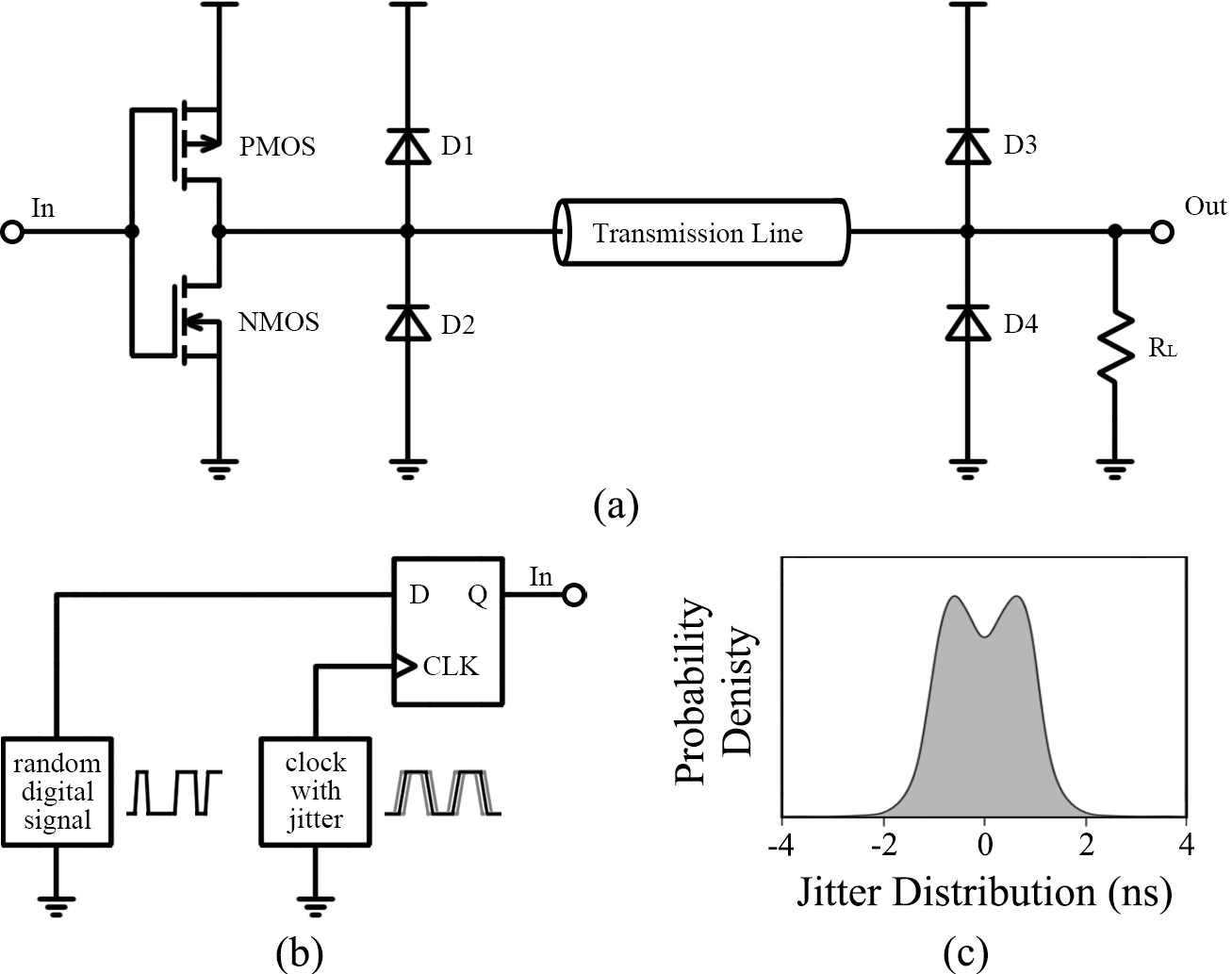}}
\caption{\textcolor{black}{SPICE models used in this paper. (a) Model of a transmission system. (b) Model of generating input signal. (c) Jitter distribution in the clock signal. }}
\label{Fig2}
\end{figure}

\begin{figure}[t]
\centerline{\includegraphics[width=180pt]{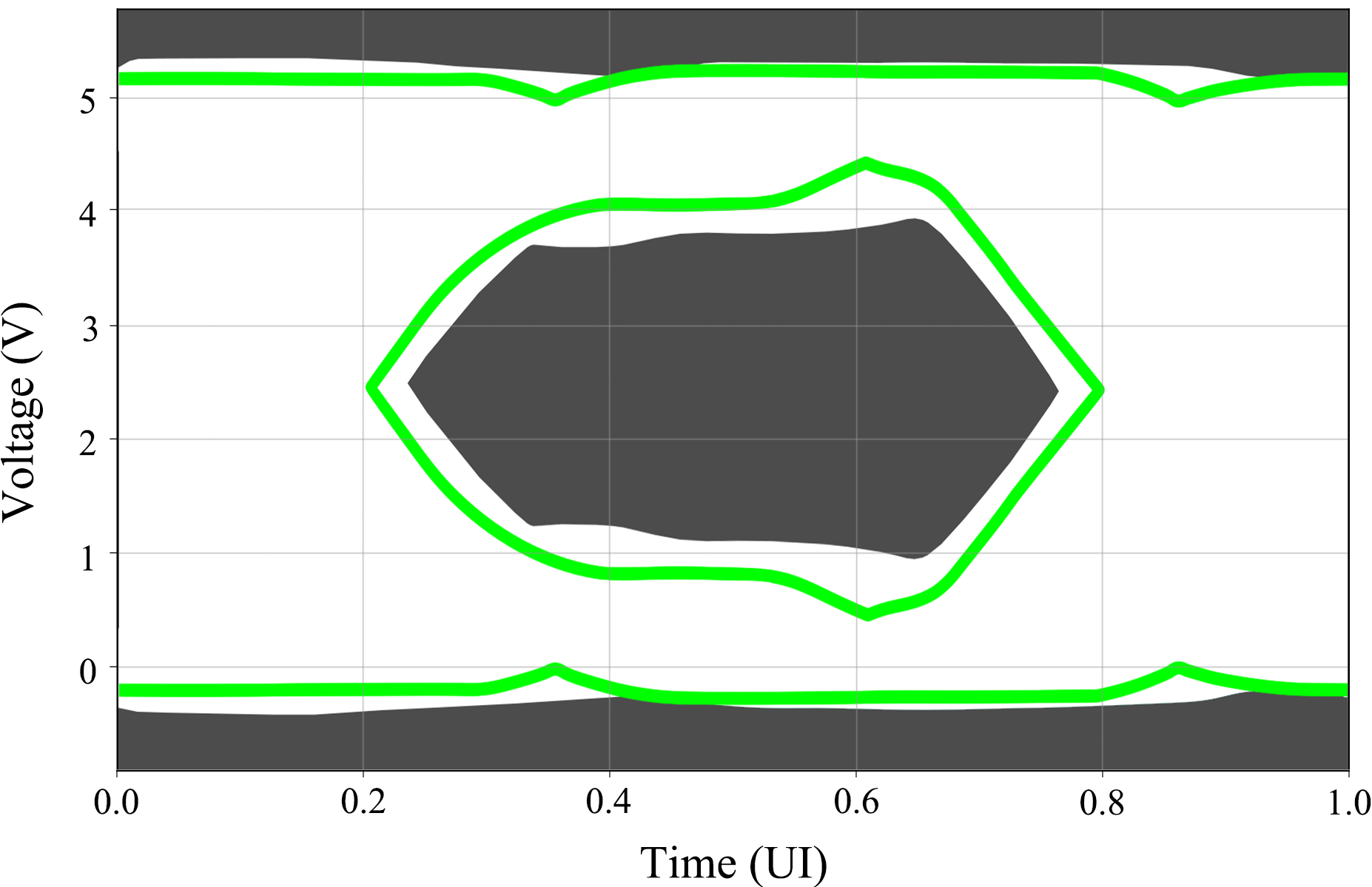}}
\caption{\textcolor{black}{Eye diagrams generated by two methods. The white area is the eye diagram formed by transient simulation. The green solid lines are the boundaries of the eye diagram formed by sub-methods in [9], [3], and [8].}}
\label{Fig3}
\end{figure}

\color{black}
For instance, we can build a model in SPICE. Input signals are input to an inverter and output signals can be detected on the output terminal as shown in Fig.~\ref{Fig2}(a). Assume the characteristics impedance of the lossy transmission line is around $50\Omega$, the load resistance is $400\Omega$, the period of the signal is $20ns$, and the jitter obeys a distribution as shown in Fig.~\ref{Fig2}(c), which contains random jitter (RJ) whose standard deviation is $\sigma_{RJ}=0.4ns$, and period jitter (PJ) whose amplitude is $A_{PJ}=1.0ns$. To illustrate the inaccuracy of the convolution process, this study represents two different methods to form an eye diagram in this section.
\begin{itemize}
\item Firstly, by using traditional transient simulation, input signals are generated by an ideal D flip-flop as shown in Fig.~\ref{Fig2}(b). The module `random digital signal' can generate 0 or 1 randomly. The module `clock with jitter' can generate clock signals in which edges are generated either early or late with certain probabilities. Both modules are behavioral voltages and their signals are represented as built-in functions.
\item Secondly, the sub-methods as detailed in the literature \cite{b9}, \cite{b3}, and \cite{b8} can also create eye diagrams. The order of MER must be established before creating eye diagrams using the sub-method \cite{b9}. Next, assume that there's no jitter in input signals so that the sub-method in \cite{b3} can form preliminary eye diagrams. Finally, use the sub-method in \cite{b8} to introduce the jitter effect. Because the probability of jitter occurrence decreases sharply as the offset of the jitter increases, this study only considers about $\pm5\sigma_{RJ}$ for the RJ.
\end{itemize}
The results using these two methods are shown in Fig.~\ref{Fig3}. (`UI' is `unit interval'.) The white area is the area of the eye diagram formed by the former method, and the green solid lines are the boundaries of the eye diagram formed by the latter method. Evidently,
\color{black}
the convolution process is not accurate. Thus, here proposes a new MER-based method.

\section{Overview of the Proposed Method}\label{II}

\color{black}
It is anticipated in MER-based methods that preceding bits and jitters may have an impact on the present bit. The impact, though, cannot last for very long. To put it another way, the impact of bits and jitters produced in the deep past may be disregarded.
\color{black}
Thus, it's necessary to determine how many previous bits and jitters should be considered. Numbers of the considered previous bits and jitters are termed orders of bit effect (BE) and jitter effect (JE) respectively.
\color{black}
In particular, the jitter's variance is less than a single full bit.
\color{black}
Thus, the order of JE should be less than the order of BE. In this paper, an algorithm to determine the orders of BE and JE is presented, as shown in Section~\ref{III}.

When these two orders are determined, we can confirm which \textcolor{black}{MERs} are required. \textcolor{black}{However, } performing \textcolor{black}{short-term} transient simulations to generate \textcolor{black}{all} these required \textcolor{black}{MERs} means a grievous waste of time. Thus, it's necessary to determine the minimal number of sampling points and then perform \textcolor{black}{short-term transient simulations to generate a very small fraction of the required MERs}. By using these \textcolor{black}{MERs}, all the required \textcolor{black}{MERs} can be restored accurately. In this paper, the proposed method \textcolor{black}{can introduce} PJ and RJ into MERs, as shown in Section~\ref{IV}.

\begin{figure}[t]
\centerline{\includegraphics[width=120pt]{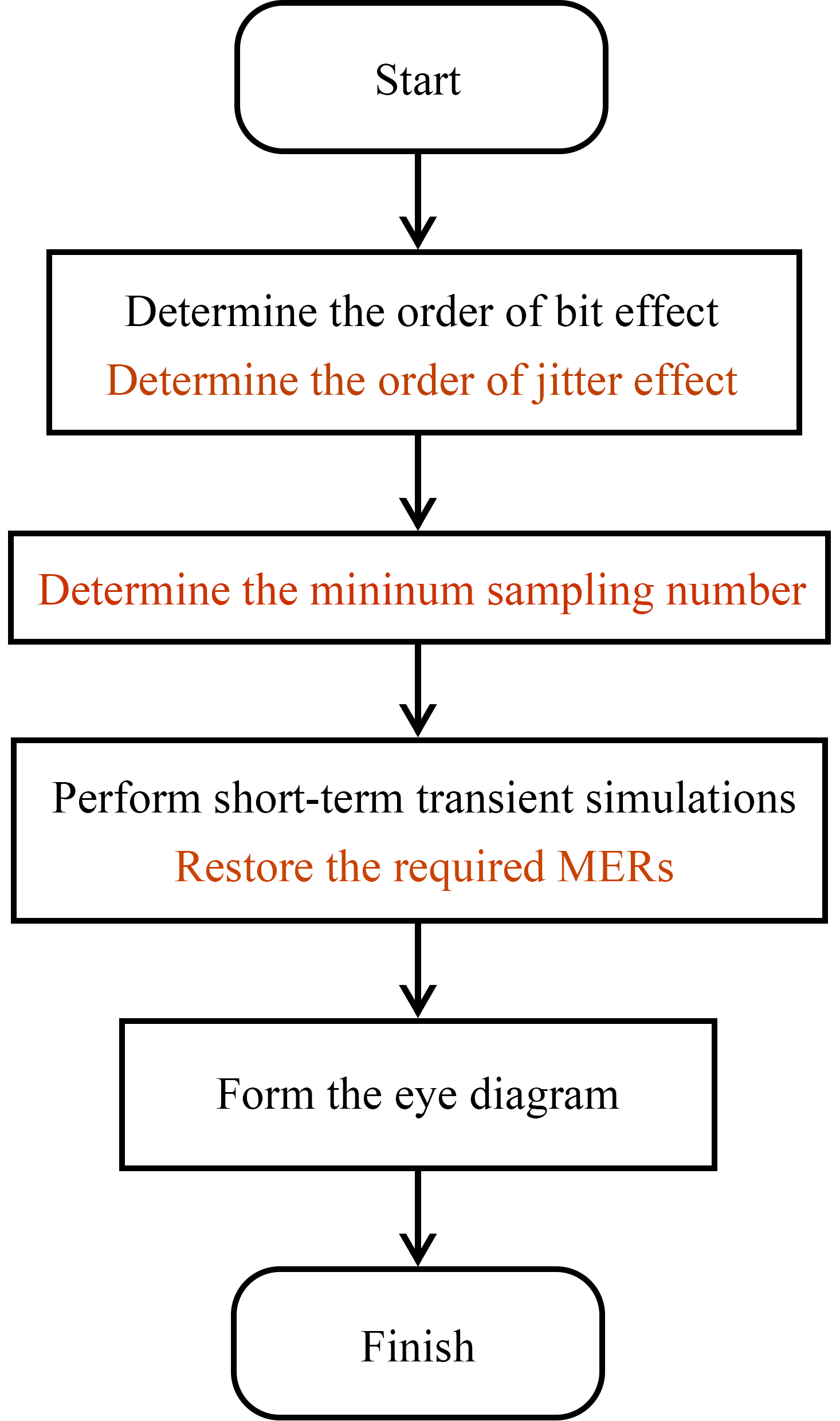}}
\caption{Process of the proposed method. \textcolor{black}{The black parts are the basic steps of MER-based methods. The brown parts are new steps for the proposed method.}}
\label{Fig4}
\end{figure}

In general, the process of the proposed method is shown in Fig.~\ref{Fig4}. The black parts are \textcolor{black}{the} basic steps of MER-based methods. The brown parts are new steps for the proposed method.

Section~\ref{V} uses \textcolor{black}{the} proposed method to estimate eye diagrams and this method is compared with transient simulation and \textcolor{black}{the previous} MER-based method. Section~\ref{VI} gives the conclusion.

\color{black}
It should be noted that the model shown in Fig. 2 is used as a reference in this paper.
\color{black}
The low and high levels are $0$ and $5V$. The characteristic impedance of the lossy transmission line is around $50\Omega$ in the frequency range.
\color{black}
In Section~\ref{III} and Section~\ref{IV}, the load resistance is $200\Omega$. In Section~\ref{V}, the load resistance is modified for verification.
\color{black}

\section{Determination of the Minimum Required Order}\label{III}

\subsection{Order of Bit Effect}\label{IIIA}

\begin{table*}[b]
\caption{Effects and Required Responses at Different Indexes}
\label{TabI}
    \begin{center}
    \begin{tabular}{|p{2.3cm}
                    |p{0.8cm}|p{0.8cm}|p{0.8cm}|p{0.8cm}|p{0.8cm}|p{0.8cm}|p{0.8cm}|p{0.8cm}
                    |p{1.25cm}|p{1.25cm}|}\hline
        \multicolumn{1}{|c|}{\textbf{index $m$}}&
            ...&
            $m_{b+1}$&
            $m_b$&
            ...&
            $m_{j+1}$&
            $m_j$&
            ...&
            $1$&
            $0$&
            $-1$\\\hline
        \multicolumn{1}{|c|}{\textbf{considered effects}}&
            \multicolumn{2}{c|}{neither}&
            \multicolumn{3}{c|}{bit effect}&
            \multicolumn{3}{c|}{bit effect and jitter effect}&
            current bit&next bit\\\hline
        \multicolumn{1}{|c|}{\textbf{whether required}}&
            \multicolumn{2}{c|}{not required}&
            \multicolumn{8}{c|}{required}\\\hline
    \end{tabular}
    \end{center}
\end{table*}

In \cite{b9}, \textcolor{black}{the authors proposed} a method to determine the order of MER. It can \textcolor{black}{commendably} evaluate how long the effect of a rising or falling edge can remain. In this paper, not only edges but also high and low levels are considered.
\color{black}
As a result, this method has to be revised.

\begin{figure}[t]
\centerline{\includegraphics[width=240pt]{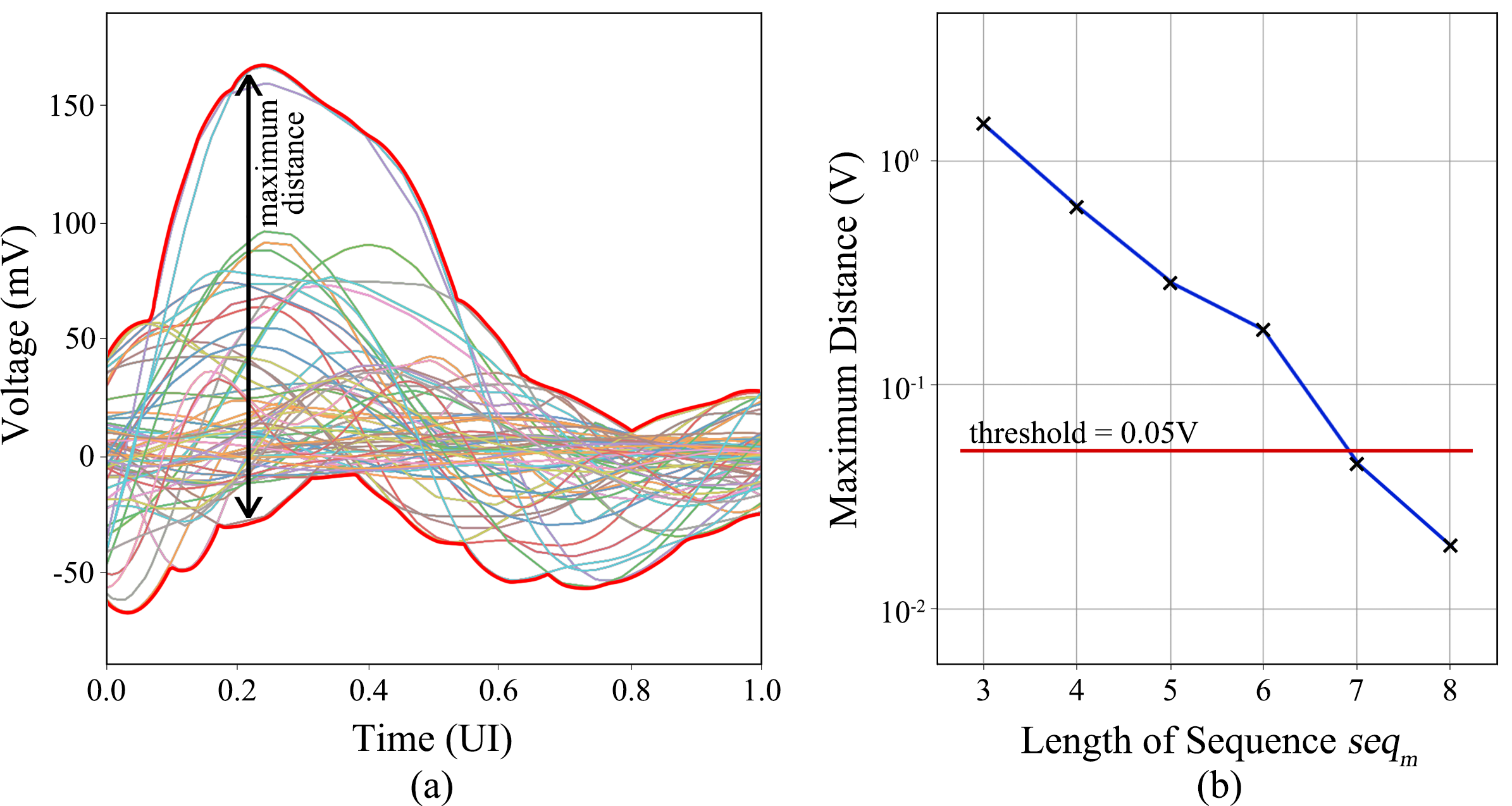}}
\caption{\textcolor{black}{Determination of the order of the bit effect. (a) Plot all the $diff_m$ on one figure when $m=6$. The upper and lower bounds are highlighted by two red lines. The maximum distance between the two bounds is highlighted by a black arrow line. (b) The maximum distance which can represent the total effect of $b_m$ decreases as the length of the input sequence $seq_m$ increases.}}
\label{Fig4-5}
\end{figure}

Suppose the input sequence is $seq_m=[b_mb_{m-1}...b_1b_0]$, bit $b_i$ generates at $t=-iT$, and the propagation time of the transmission system is $T_{delay}$. When $-iT+T_{delay}<t<-(i-1)T+T_{delay}$, the response of bit $b_i$ propagates to the receiver terminal. Because the response of bit $b_0$ is required, for each $seq_m$, record the signal as $S(seq_m)$ at the receiver terminal when $T_{delay}<t<T+T_{delay}$. Moreover, because of the effect of bit $b_m$, when $b_m$ is equal to $0$ or $1$, $S(seq_m)$ is different. Thus, the effect of bit $b_m$ in $seq_m$ can be represented as
\color{black}
\begin{equation}
diff_m=S(seq_m(b_m=0))-S(seq_m(b_m=1))
\label{eq1}
\end{equation}
\color{black}
$diff_m$ will be altered if one or more of $b_i$ ($m-1\leq i\leq0$) are changed. Thus, $seq_m$ should be changed to get different corresponding $diff_m$.
\color{black}
If $m$ is too big to simulate all the possibilities, it is possible to generate $seq_m$ randomly and simulate for fewer times relatively. Then, plot these $diff_m$ onto one figure, as shown in Fig.~\ref{Fig4-5}(a). The \textcolor{black}{total} effect of $b_m$ is reflected \textcolor{black}{in} the \textcolor{black}{maximum} distance between the upper and lower bounds \textcolor{black}{along the voltage axis direction}.
\color{black}
The bounds and the maximum distance have been highlighted respectively by two red lines and a black arrow line as shown in Fig.~\ref{Fig4-5}(a).

By altering the value of $m$, the corresponding maximum distances changes. Fig.~\ref{Fig4-5}(b) shows the curve of maximum distance against $m$,
\color{black}
where the threshold is $1\%$ of the signal amplitude (can be adjusted as needed). Thus, in the current case, the order of BE is $6$.

\subsection{Order of Jitter Effect}\label{IIIB}

\begin{figure}[t]
\centerline{\includegraphics[width=240pt]{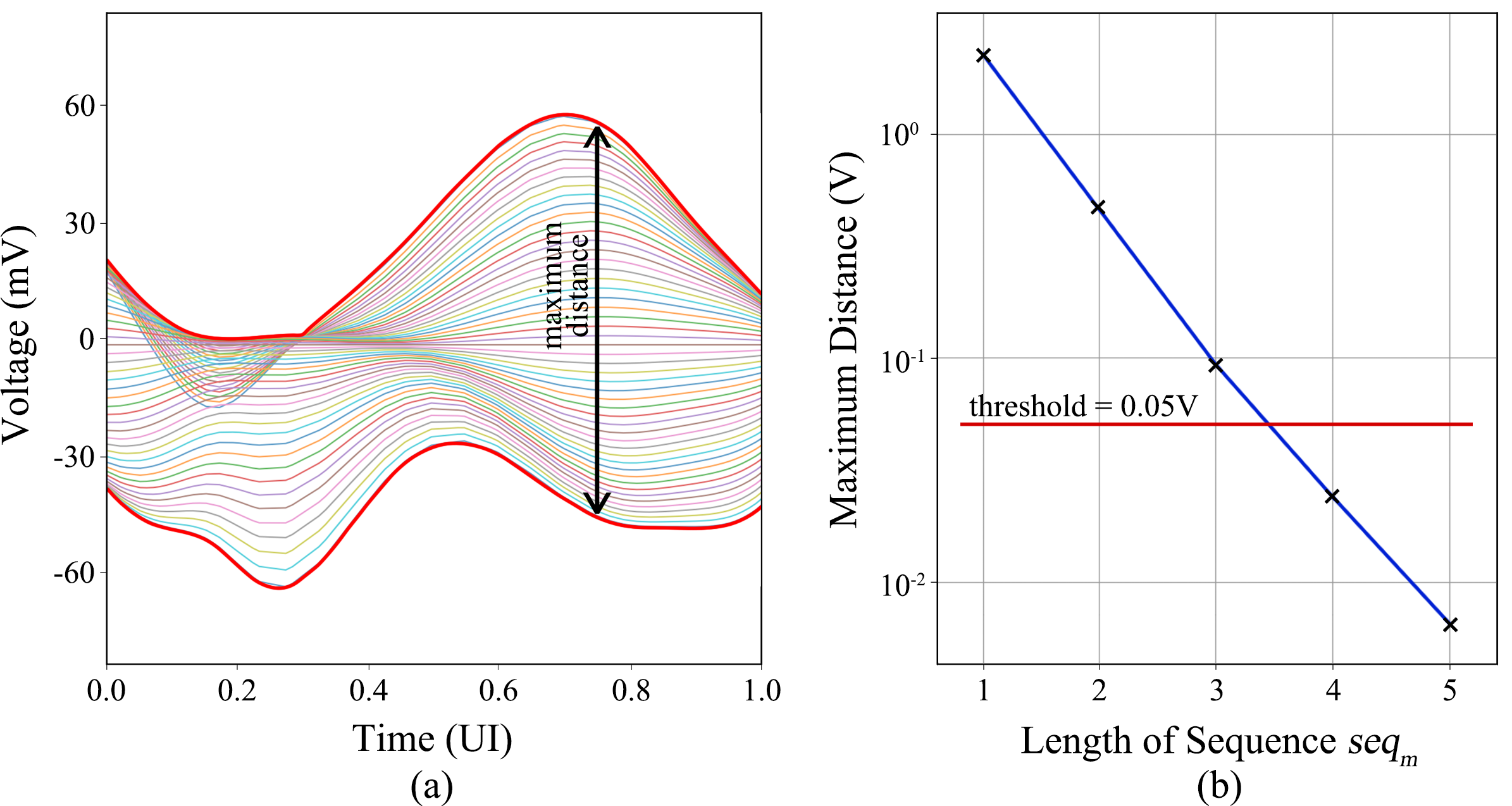}}
\caption{\textcolor{black}{Determination of the order of the jitter effect. (a) Plot all the $diff_m$ on one figure when $m=3$. The upper and lower bounds are highlighted by two red lines. The maximum distance between the two bounds is highlighted by a black arrow line. (b) The maximum distance which can represent the total effect of the jitter between bits $b_{m+1}$ and $b_m$ decreases as the length of the input sequence $seq_m$ increases.}}
\label{Fig6-7}
\end{figure}

\color{black}
The method to determine the order of JE is similar to that of BE. The order of JE represents whether a jitter between two previous bits can affect the current bit. Suppose the input sequence is $seq_m=[b_{m+1}b_mb_{m-1}...b_1b_0]$ where $b_{m+1}\neq b_m$, bit $b_i$ generates at $t=-iT$ (excluding $i=m$), and bit $b_m$ generates at $t=-mT+T_{jit}$ where $T_{jit}$ is the value of jitter between bits $b_{m+1}$ and $b_m$.
\color{black}
$T_{delay}$ is the same as the former definition.
\color{black}
As the same as BE, the output signal $S(seq_m)$ is recorded when $T_{delay}<t<T+T_{delay}$. Thus, the effect of jitter between $b_{m+1}$ and $b_m$ can be represented as
\color{black}
\begin{equation}
diff_m=S(seq_m(T_{jit}=T_x))-S(seq_m(T_{jit}=0))
\label{eq2}
\end{equation}
If the minimum and maximum values of $T_{jit}$ are $T_{min}$ and $T_{max}$, \textcolor{black}{we should} change $T_x$ from $T_{min}$ to $T_{max}$, perform \textcolor{black}{short-term transient simulations}, and record the corresponding $diff_m$. Then, these $diff_m$ are plotted into one figure, as shown in Fig.~\ref{Fig6-7}(a). The remaining process is similar to determining the order of BE. The graph of maximum distance against $m$ is shown in Fig.~\ref{Fig6-7}(b), where the threshold is $1\%$ of the signal amplitude. Thus, in the current case, the order of JE is $3$.

\subsection{Required Responses}\label{IIIC}

\color{black}
After establishing these two orders, it can be chosen which bits and jitters need to be considered.
\color{black}
Assuming that the order of BE is $m_b$ and the order of JE is $m_j$, the required bits and jitters are shown in Table.~\ref{TabI}.

\section{Generation of Required Responses}\label{IV}

\subsection{Jitter Categories for Consideration}\label{IVA}

\color{black}
This paper considers PJ and RJ.
\color{black}
In general, PJ can be represented as
\begin{equation}
\delta t_{PJ}=A_{PJ}\ sin(\frac{2\pi}{T_{PJ}}t)
\label{eq3}
\end{equation}
where $\delta t_{PJ}$ is the offset caused by PJ. $A_{PJ}$ and $T_{PJ}$ are the amplitude and the period of PJ, respectively. RJ is subject to Gaussian distribution, which is represented as
\begin{equation}
\delta t_{RJ}\sim N(0,\sigma_{RJ}^2)=\frac{1}{\sqrt{2\pi}\sigma_{RJ}}
exp(-\frac{\delta t_{RJ}^2}{2\sigma_{RJ}^2})
\label{eq4}
\end{equation}
where $\delta t_{RJ}$ is the offset caused by RJ. $\sigma_{RJ}$ is the standard deviation of RJ. In an $m_j$-order JE system, suppose there is an edge between bits $m$ and $m+1$ ($m\leq m_j$). The moment of bit $m$ generating is $t=-mT+\delta t_{PJ}+\delta t_{RJ}$.

\subsection{Method of \textcolor{black}{Introducing} Period Jitter Into Signals}\label{IVB}

First, take PJ into account. The moment of bit $m$ generating is
\begin{equation}
t=-mT+A_{PJ}\ sin(\frac{2\pi}{T_{PJ}}(-t+T_0))
\label{eq5}
\end{equation}
where $A_{PJ}$ is small and $t$ is close to $-mT$. \textcolor{black}{To} simplify the calculation, the moment of bit $m$ generating is approximately \textcolor{black}{equal} to
\begin{equation}
t=-mT+A_{PJ}\ sin(\frac{2\pi}{T_{PJ}}(-mT+T_0))
\label{eq6}
\end{equation}
Thus, the input signal can be represented as
\begin{equation}
\begin{split}
si&gnal(t)=\ level\\
+\sum_{m=m_b}^{m_{j+1}}(b_m-b_{m+1})&U(t+mT)+\sum_{m=m_j}^{-1}(b_m-b_{m+1})\\
U(t+mT-&A_{PJ}\ sin(\frac{2\pi}{T_{PJ}}(-mT+T_0)))
\end{split}
\label{eq7}
\end{equation}
where $U(t)$ is the edge sample at \textcolor{black}{the} input terminal and $level$ is the initial level. To iterate through all the possibilities, $T_0$ must get the value from $0.01T_{PJ}$ to $T_{PJ}$.
\color{black}
Suppose the step size is $0.01T_{PJ}$.
\color{black}
In an $m_b$-order BE system,
\color{black}
the number of required MERs is $2^{m_b+2}(T_{PJ}/0.01T_{PJ})$. It will require a lot of computation if solely employing short-term transient simulations to obtain these MERs.
To accelerate the process, this method aims to get a small portion of MERs by performing short-term transient simulations and use these MERs to restore the others. Thus, it's necessary to figure out the minimal number of sampling points.
\color{black}

\begin{figure}[t]
\centerline{\includegraphics[width=250pt]{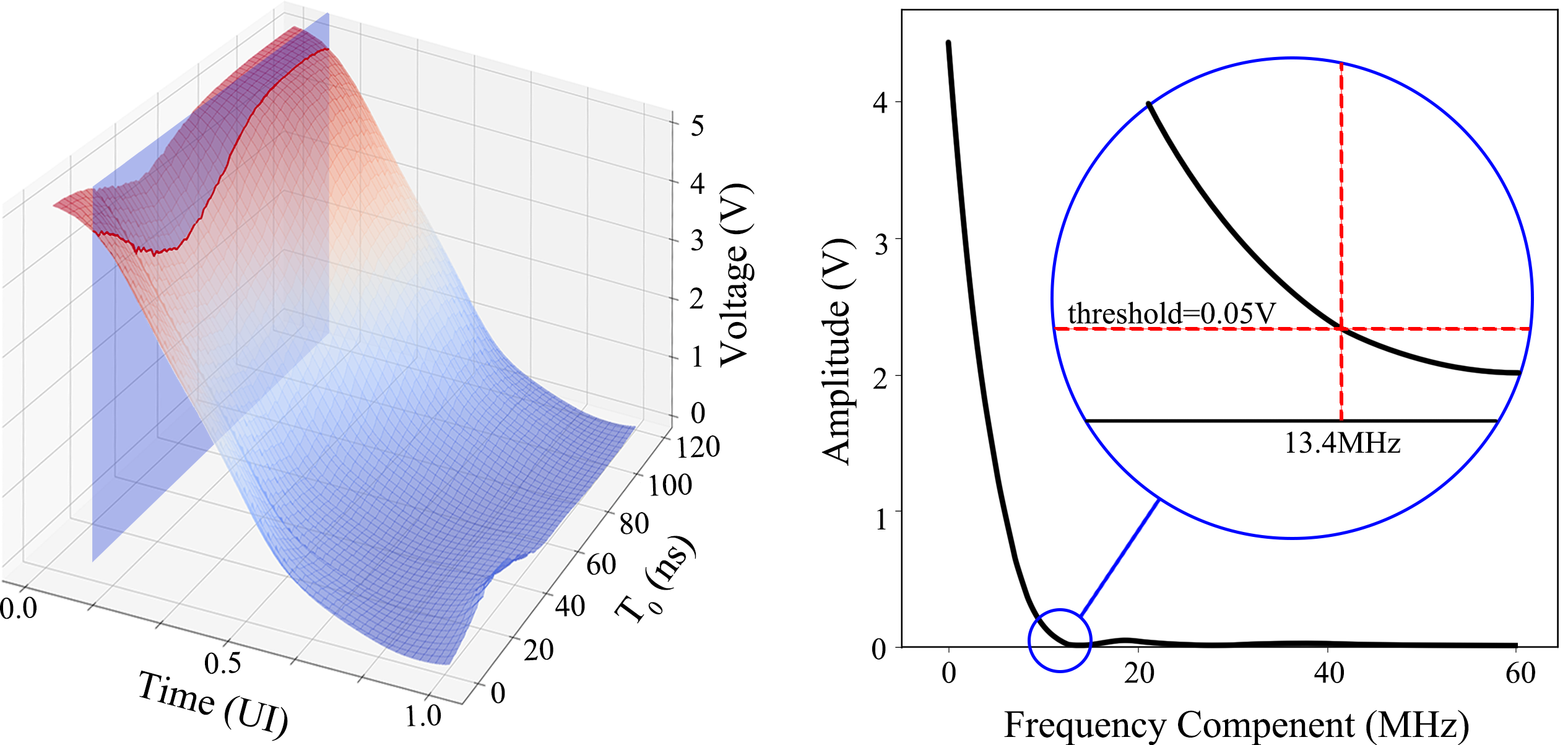}}
\caption{\textcolor{black}{(a) The 3-dimensional surface formed by all the output signals $S(T_0)$ which change as $T_0$ changes. Another surface paralleling to the plane $T_0-Voltage$ slices the original surface. The sliced curve reflects the effect of $T_0$ at a certain moment. (b) Spectrum of the curve generated by slicing the 3-dimensional figure. Frequency components whose amplitudes are lower than the threshold can be omitted.}}
\label{Fig8-9}
\end{figure}

\begin{figure}[t]
\centerline{\includegraphics[width=160pt]{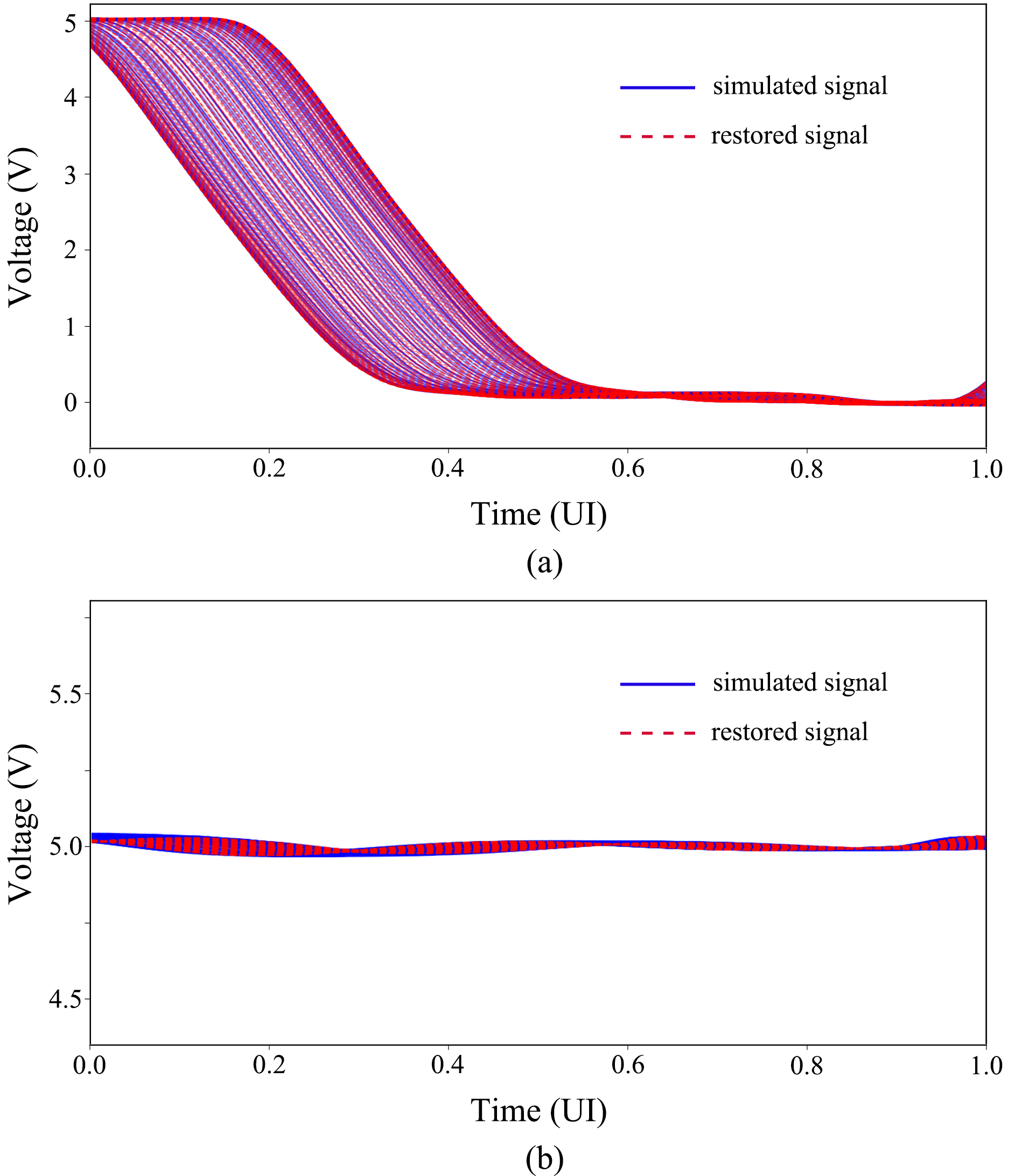}}
\caption{\textcolor{black}{Signals generated by two methods while considering the period jitter. Blue solid lines are generated by transient simulation. Red dot lines are generated by sampling and restoring. (a) When $seq=[0101010]$. (b) When $seq=[1101001]$.}}
\label{Fig11}
\end{figure}

In an \textcolor{black}{$m_b$-order BE and} $m_j$-order JE system, suppose \textcolor{black}{the input sequence is $seq=[...01010]$, so the signal can be represented as}
\begin{equation}
\begin{split}
signa&l(t)=\ level\\
+\sum_{m=m_b}^{m_{j+1}}(-1)^mU(&t+mT)+\sum_{m=m_j}^{-1}(-1)^m\\
U(t+mT-A_{PJ}&\ sin(\frac{2\pi}{T_{PJ}}(-mT+T_0)))
\end{split}
\label{eq8}
\end{equation}
where $T_0$ runs from $0.01T_{PJ}$ to $T_{PJ}$. \textcolor{black}{This method first performs short-term} transient simulations for $T_{PJ}/0.01T_{PJ}=100$ times and \textcolor{black}{records} output \textcolor{black}{signals} $S(T_0)$ when $T_{delay}<t<T+T_{delay}$.
\color{black}
Then, it plots all the $S(T_0)$ onto one figure to form a 3-dimensional surface, and slices this surface by another surface paralleling to the plane $T_0-Voltage$, as shown in Fig.~\ref{Fig8-9}(a). The intersection curve of two surfaces reflects the effect of $T_0$ at a certain moment $Time$. Next, the spectrum of this curve can be calculated by performing Fast Fourier Transform (FFT), where high-frequency components are not evident so that they can be omitted for acceleration. Thus, a threshold is set to figure out the cut-off frequency, as shown in Fig.~\ref{Fig8-9}(b).
\color{black}
For instance, in the current case, the threshold is $1\%$ of the signal amplitude, \textcolor{black}{so} the cut-off frequency is \textcolor{black}{$f_{cut}=13.4MHz$}.
\color{black}
Next, the surface paralleling to the plane $T_0-Voltage$ should be shifted to get the intersection curve and calculate all the cut-off frequencies $f_{cut}$. The maximum one is regarded as the required cut-off frequency $f_{cut,max}$. According to the Nyquist Theory, the minimum sampling frequency $f_s$ needs to meet $f_s>2f_{cut,max}$. In the current case, the required cut-off frequency is $f_{cut,max}=14.6MHz$, and the sampling frequency $f_s$ is equal to $3f_{cut,max}=43.8MHz$. Thus, the sampling time is $T_s=1/f_s=22.8ns$. In the current case, the period of PJ is $T_{PJ}=118ns$. Thus, simulating for $118ns/22.8ns=5.17\approx6$ times, all the MERs can be restored. To validate the method, these restored and simulated MERs are plotted into one figure, as shown in Fig.~\ref{Fig11}(a), which is exact.
\color{black}

Moreover, \textcolor{black}{MERs} of other \textcolor{black}{input sequences} $seq$ can be restored by using the same \textcolor{black}{sampling frequency} $f_s$. For instance, the restored \textcolor{black}{MERs} and the simulated \textcolor{black}{MERs} of sequence $seq=[1101001]$ are shown in Fig.~\ref{Fig11}(b).

\subsection{Method of \textcolor{black}{Introducing} Random Jitter Into Signals}\label{IVC}

The method to take RJ into account is similar to PJ. The input signal can be represented as
\begin{equation}
\begin{split}
si&gnal(t)=\ level+\\
\sum_{m=m_b}^{m_{j+1}}(b_m-b_{m+1})&U(t+mT)+\sum_{m=m_j}^{-1}(b_m-b_{m+1})\\
U(t+mT-A_{PJ}&\ sin(\frac{2\pi}{T_{PJ}}(-mT+T_0))-\delta t_{RJ m})
\end{split}
\label{eq10}
\end{equation}
Suppose \textcolor{black}{the input sequence $seq$ and} $T_0$ are fixed and we need to change each $\delta t_{RJ m}$, which runs from $-5\sigma_{RJ}$ to $+5\sigma_{RJ}$ (can be changed as needed). Suppose the step size is $0.1\sigma_{RJ}$. To iterate through all the possibilities,
\color{black}
there are $(10\sigma_{RJ}/0.1\sigma_{RJ})^{m_j+2}=100^{m_j+2}$ required MERs. It is horrible to get them by performing short-term transient simulations solely. Thus, it's necessary to use the same method as Section~\ref{IVB} for acceleration.
\color{black}

\begin{table}[tb]
\caption{$f_{cut}$, $f_s$, $T_s$, and $num_m$ at Different Indexes}
\label{TabIII}
    \begin{center}
    \begin{tabular}{|p{2.0cm}|p{0.8cm}|p{0.8cm}|p{0.8cm}|p{0.8cm}|p{0.8cm}|}\hline
        \multicolumn{1}{|c|}{\textbf{index $m$}}
            &$3$        &$2$        &$1$        &$0$        &$-1$\\\hline
        \multicolumn{1}{|c|}{\textbf{$f_{cut.max}/GHz$}}
            &$0.125$    &$0.208$    &$0.613$    &$2.08$     &$0.820$\\\hline
        \multicolumn{1}{|c|}{\textbf{$f_s/GHz$}}
            &$0.375$    &$0.624$    &$1.84$     &$6.24$     &$2.46$\\\hline
        \multicolumn{1}{|c|}{\textbf{$T_s/ns$}}
            &$2.67$     &$1.60$     &$0.543$    &$0.160$    &$0.406$\\\hline
        \multicolumn{1}{|c|}{\textbf{$num_m$}}
            &$1$        &$2$        &$5$        &$15$        &$6$\\\hline
    \end{tabular}
    \end{center}
\end{table}

In an \textcolor{black}{$m_b$-order BE and} $m_j$-order JE system, suppose the input sequence is \textcolor{black}{$seq=[...01010]$, so the signal} can be represented as
\begin{equation}
\begin{split}
signa&l(t)=\ level\\
+\sum_{m=m_b}^{m_{j+1}}(-1)^mU(&t+mT)+\sum_{m=m_j}^{-1}(-1)^m\\
U(t+mT-A_{PJ}\ sin(&\frac{2\pi}{T_{PJ}}(-mT+T_0))-\delta t_{RJ m})
\end{split}
\label{eq11}
\end{equation}
There are $(m_j+2)\ \delta t_{RJ m}$ in this formula. \textcolor{black}{This method makes one} $\delta t_{RJ k}$ ($-1\leq k\leq m_j$) \textcolor{black}{vary} from $-5\sigma_{RJ}$ to $+5\sigma_{RJ}$ and makes the others fixed. Then, \textcolor{black}{it performs} the same process as Section~\ref{IVB}.
\color{black}
It plots all the $S(\delta t_{RJ k})$ into one figure to form a surface, slices this surface to get intersection curves and their spectrums, and sets a threshold to get the maximum cut-off frequency $f_{cut,max}$. Finally, the minimal number of sampling points $num_m$ at the certain index $m$ is figured out.
\color{black}
The corresponding \textcolor{black}{$f_{cut,max}$}, $f_s$ ($3$ times $f_{cut}$), $T_s$, and the number of sampling points $num_m$ have been shown in Table.~\ref{TabIII}. Especially, if the number of sampling points is $1$, it means that, for the range of RJ at the certain index \textcolor{black}{$m$}, these $S(\delta t_{RJ k})$ are almost unaffected by RJ and one single sample is enough.

\begin{figure}[tb]
\centerline{\includegraphics[width=160pt]{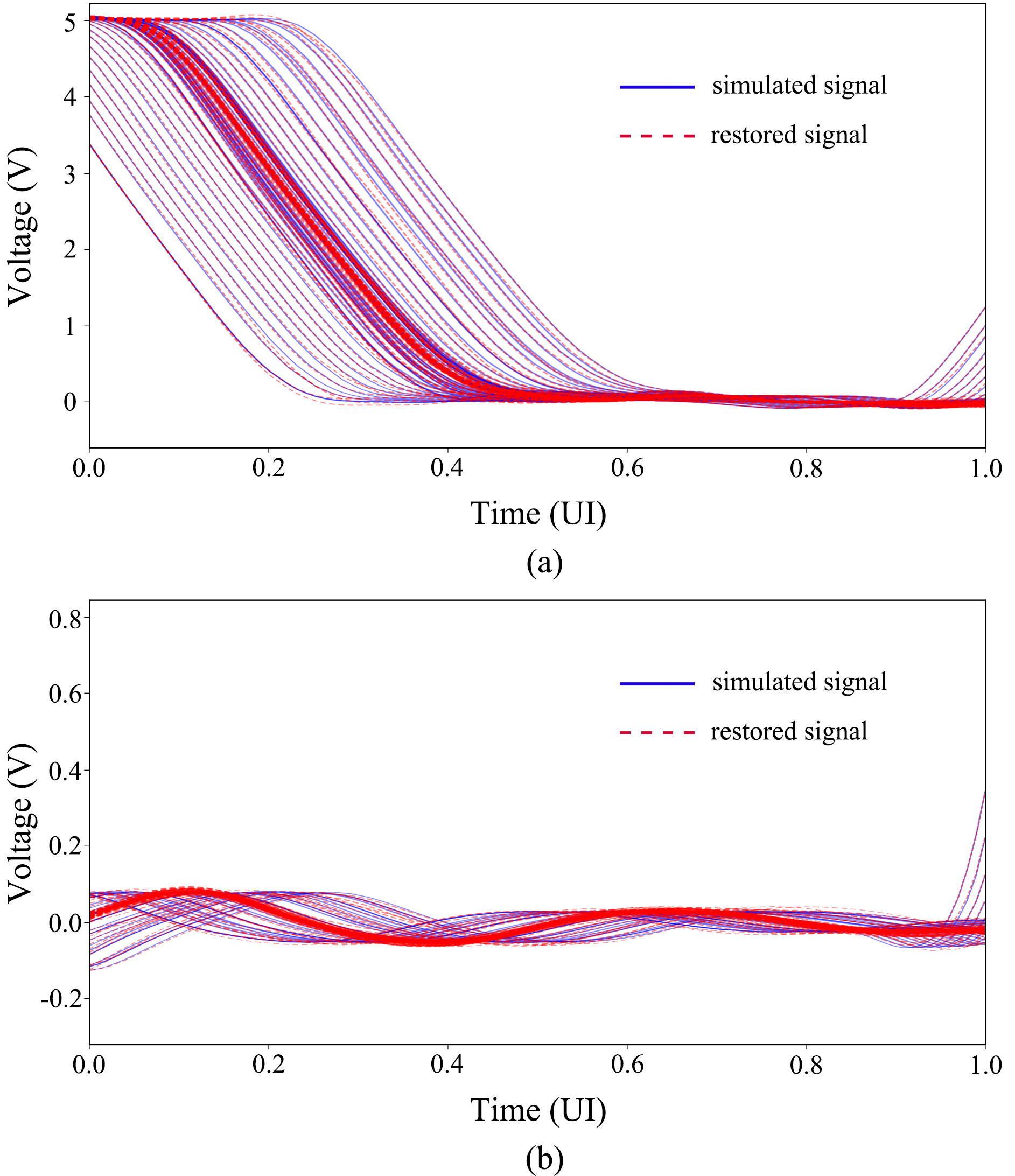}}
\caption{\textcolor{black}{Signals generated by two methods while considering the random jitter. Blue solid lines are generated by transient simulation. Red dot lines are generated by sampling and restoring. (a) When $seq=[0101010]$. (b) When $seq=[1001110]$.}}
\label{Fig12}
\end{figure}

\begin{table}[t]
\caption{Order of MER Calculated by Two Methods}
\label{TabIV}
    \begin{center}
    \begin{tabular}{|p{2.62cm}|p{0.45cm}|p{0.45cm}|p{0.45cm}|p{0.45cm}|p{0.45cm}|p{0.45cm}|}\hline
        \multicolumn{2}{|c|}{\textbf{load resistance} / $\Omega$}
            & $10$ & $20$ & $50$ & $200$ & $400$ \\ \hline
        \multirow{2}{*}{\textbf{the proposed method}}&
            \multicolumn{1}{|c|}{JE}
             & $4$ & $3$ & $2$ & $3$ & $5$ \\ \cline{2-7}
        \multicolumn{1}{|c|}{}&
            \multicolumn{1}{|c|}{BE}
            & $7$ & $5$ & $4$ & $6$ & $9$ \\ \hline
        \multicolumn{2}{|c|}{\textbf{the method in [3]}}
            & $7$ & $5$ & $4$ & $6$ & $8$ \\ \hline
    \end{tabular}
    \end{center}
\end{table}

To restore all the \textcolor{black}{MERs} at moment $Time$, we can construct an m-dimensional matrix \textcolor{black}{whose} size is $num_{-1}\times num_0\times ...\times num_m$. Then, \textcolor{black}{it can be restored} one dimension by one dimension. For instance, the original $num_{-1}\times num_0\times ...\times num_m$ matrix can be restored as $100\times num_0\times ...\times num_m$, then $100\times 100\times ...\times num_m$, and so on. Finally, it can be transformed to $100\times 100\times ...\times 100$. Fig.~\ref{Fig12}(a) shows some of the restored \textcolor{black}{MERs} and the simulated \textcolor{black}{MERs}. The result is accurate.

Moreover, \textcolor{black}{MERs} of other \textcolor{black}{input sequences} $seq$ can be restored by using the same $f_s$. For instance, the restored \textcolor{black}{MERs} and the simulated \textcolor{black}{MERs} of sequence $seq=[1001110]$ are shown in Fig.~\ref{Fig12}(b).

\section{Verification}\label{V}

\begin{table*}[t]
\caption{Eye Heights and Eye Widths Calculated by Three Methods}
\label{TabV}
    \begin{center}
    \begin{tabular}{|p{1.8cm}|p{2.7cm}|p{1.6cm}|p{1.6cm}|p{1.6cm}|p{1.6cm}|p{1.6cm}|}\hline
        \multicolumn{2}{|c|}{\textbf{load resistance} / $\Omega$}
            & $10$
            & $20$
            & $50$
            & $200$
            & $400$ \\ \hline
        \multirow{3}{*}{\textbf{eye height} / $V$} &
        \multicolumn{1}{|c|}{\textbf{transient simulation}}
            & $0.922$
            & $2.00$
            & $4.32$
            & $3.88$
            & $2.82$ \\ \cline{2-7}
        \multicolumn{1}{|c|}{} &
        \multicolumn{1}{|c|}{\textbf{the prior method}}
            & $1.74\ (89\%)$
            & $2.56\ (28\%)$
            & $4.48\ (3.7\%)$
            & $4.11\ (5.9\%)$
            & $3.19\ (13\%)$ \\ \cline{2-7}
        \multicolumn{1}{|c|}{} &
        \multicolumn{1}{|c|}{\textbf{the proposed method}}
            & $0.980\ (6.3\%)$
            & $1.95\ (2.5\%)$
            & $4.38\ (1.4\%)$
            & $3.92\ (1.0\%)$
            & $2.74\ (2.8\%)$ \\ \hline
        \multirow{3}{*}{\textbf{eye width} / $UI$} &
        \multicolumn{1}{|c|}{\textbf{transient simulation}}
            & $0.354$
            & $0.452$
            & $0.574$
            & $0.550$
            & $0.531$ \\ \cline{2-7}
        \multicolumn{1}{|c|}{} &
        \multicolumn{1}{|c|}{\textbf{the prior method}}
            & $0.456\ (29\%)$
            & $0.507\ (12\%)$
            & $0.600\ (4.5\%)$
            & $0.589\ (7.1\%)$
            & $0.587\ (11\%)$ \\ \cline{2-7}
        \multicolumn{1}{|c|}{} &
        \multicolumn{1}{|c|}{\textbf{the proposed method}}
            & $0.367\ (3.7\%)$
            & $0.445\ (1.5\%)$
            & $0.580\ (1.0\%)$
            & $0.548\ (0.36\%)$
            & $0.538\ (1.3\%)$ \\ \hline
        \multicolumn{7}{l}{In the rows of \textbf{the prior method} and \textbf{the proposed method}, the numbers in parentheses are the errors with respect to \textbf{transient}} \\
        \multicolumn{7}{l}{\textbf{simulation}. For example, when the load resistance is $10\Omega$, the error of \textbf{the prior method} is $(1.74-0.922)/0.922=89\%$.}
    \end{tabular}
    \end{center}
\end{table*}

\begin{figure*}[t]
\centerline{\includegraphics[width=460pt]{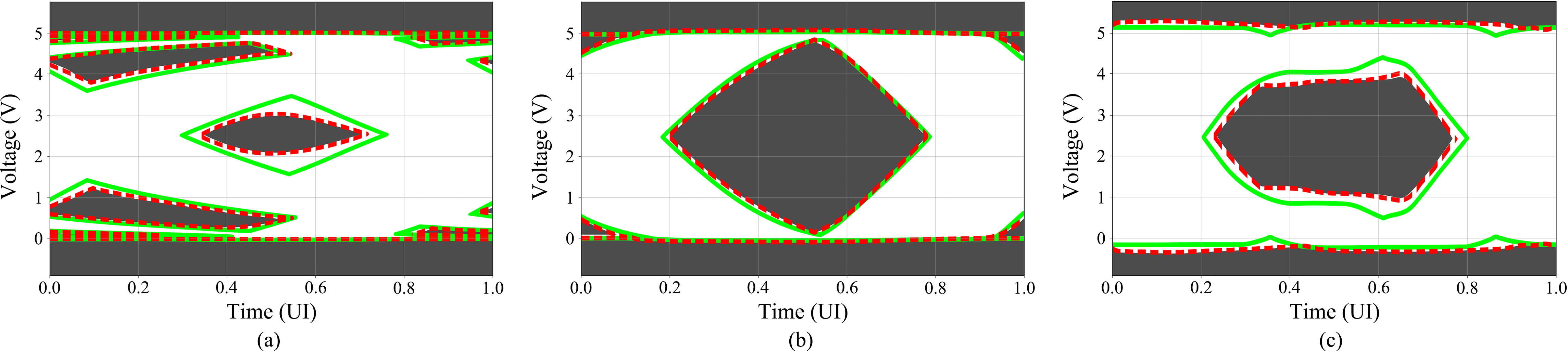}}
\caption{\textcolor{black}{Eye diagrams generated by three methods. The white areas are eye diagrams formed by transient simulation. The green solid lines are the boundaries of eye diagrams formed by the prior method (sub-methods in [9], [3], and [8]). The red dot lines are the boundaries of eye diagrams formed by the proposed method.} (a) \textcolor{black}{When the} load resistance=$10\Omega$. (b) \textcolor{black}{When the} load resistance=$50\Omega$. (c) \textcolor{black}{When the} load resistance=$400\Omega$.}
\label{Fig13}
\end{figure*}

By using the proposed method, we can estimate the eye diagram as the case shown in Fig.~\ref{Fig2}. Suppose the period of the input signal is $T=20ns$, the amplitude and the period of PJ are $A_{PJ}=1.0ns$ and $T_{PJ}=118ns$, and the standard deviation of RJ is $\sigma_{RJ}=0.4ns$. When the load resistance is different from the characteristic impedance (around $50\Omega$) of the transmission line, the quality of the output signal will get bad. In this instance, the load resistance is set as $10\Omega$, \textcolor{black}{$20\Omega$}, $50\Omega$, \textcolor{black}{$200\Omega$}, and $400\Omega$.

There are three methods to be compared:
\color{black}
two methods mentioned in Section~\ref{I} and the proposed method. First, this study determines orders of MER, which are shown in Table.~\ref{TabIV}. In most instances, the order of BE calculated by the proposed method is equal to the order of MER calculated by the method in [9]. In some other instances, they have minor differences. According to these orders, eye diagrams can be created. Considering the results of transient simulation as standards, values and errors of eye heights and eye widths calculated by the other two methods are shown in Table.~\ref{TabV}. The formed eye diagrams when the load resistances are $10\Omega$, $50\Omega$, and $400\Omega$ are shown in Fig.~\ref{Fig13}. White areas are eye diagrams formed by transient simulation, green solid lines are the boundaries of eye diagrams formed by the previous method, and red dotted lines are the boundaries of eye diagrams formed by the proposed method.

As the result shows,
\color{black}
when the load resistance is much different from the characteristic impedance, the result of \textcolor{black}{the} convolution process \textcolor{black}{becomes} imprecise but the proposed method is much better.

\section{Conclusion}\label{VI}

\color{black}
This paper has presented an enhanced MER-based method that can estimate eye diagrams accurately. First, it has detailed the algorithm for determining the orders of BE and JE. After that, it has detailed and validated the method to calculate the minimum sampling number and introduce PJ and RJ into MERs. Finally, the results generated by the proposed method have been compared with two other methods. It has indicated that
\color{black}
the more severe the mismatch between the transmission line and the load resistance, the more superior the proposed method.


\begin{figure*}[t]
\centerline{\includegraphics[width=600pt]{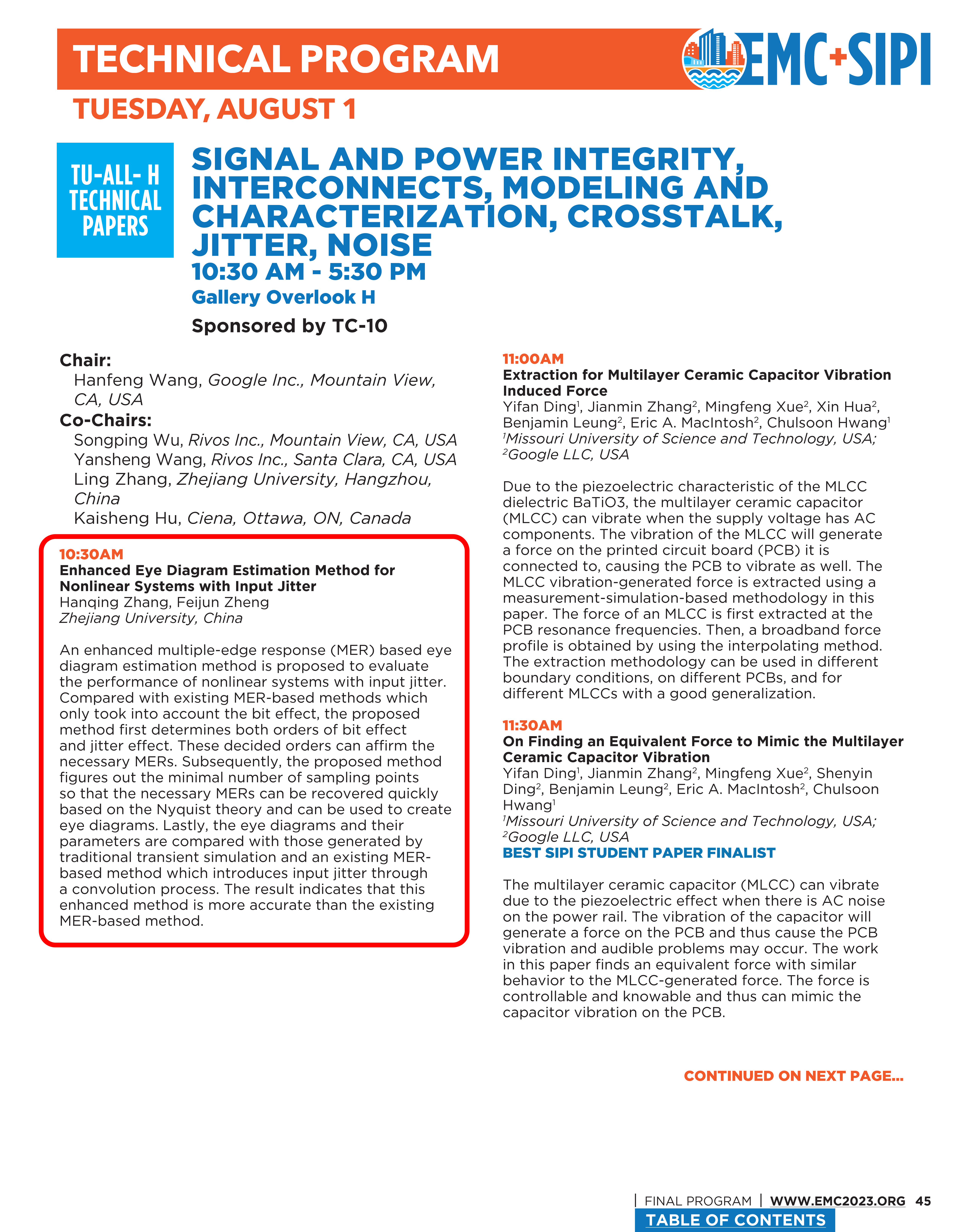}}
\end{figure*}

\end{document}